\documentclass[9pt, twocolumn, letterpaper]{./IEEEtran}
\RequirePackage{times}

\RequirePackage[usenames]{pstricks} 
\RequirePackage{graphicx}
\RequirePackage{rotating}
\RequirePackage{indentfirst} 
\RequirePackage{wrapfig}
\RequirePackage{color} 
\RequirePackage{epsfig}
\RequirePackage{eepic} 
\RequirePackage{subfigure}
\RequirePackage{amsbsy} 
\RequirePackage{amsmath}
\RequirePackage{amssymb}
\RequirePackage{latexsym}
\RequirePackage{mathptm} 
\DeclareMathVersion{bold} 
\RequirePackage[final,notref]{showkeys} 
\RequirePackage{bm}
\RequirePackage{multicol}
\RequirePackage{esvect} 
\RequirePackage{ifthen}
\usepackage{setspace}
\usepackage{moreverb}
\usepackage{ulem}
\usepackage{mathptm}
\usepackage{amsmath}
\usepackage{mathptmx}
\usepackage{amsfonts}
\usepackage{dotlessj} 
\usepackage[sectionbib]{chapterbib}

\newcommand{\ifthen}[2]{\ifthenelse{#1}{#2}{}}

	\RequirePackage[hypertex]{hyperref}

\newcommand{\ignore}[1]{}

\newcommand{\COMMENT}[1]{}
\newcommand{\ddt}[2]{{\frac{d}{d #2}\left. #1 \right.}}

\renewcommand{\jmath}{j}

\newcommand{\halmos}{\hskip\textwidth minus\textwidth \rule{6pt}{6pt}}

\newcommand{\Real}{\Bbb{R}}

\newcommand{\Jstarphi}{J_{\phi}^*}
\newcommand{\vecbphi}{\vec b_{\phi}}
\newcommand{\vecbext}{\vec b_{\text{ext}}}
\newcommand{\phistarIdeal}{\phi_{\text{ideal}}^*}

\newtheorem{lemma}{Lemma}[section]
\newtheorem{assumption}{Assumption}[section]

\newtheorem{theorem}{Theorem}

\newcommand{\assref}[1]{Assumption \ref{ass:#1}}
\newcommand{\asslabel}[1]{\label{ass:#1}}
\newcommand{\eqnref}[1]{(\ref{eq:#1})}

\newcommand{\eqnlabel}[1]{\label{eq:#1}}
\newcommand{\figlabel}[1]{\label{fig:#1}}
\newcommand{\figref}[1]{Figure~\ref{fig:#1}}

\newcommand{\seclabel}[1]{\label{sec:#1}}
\newcommand{\secref}[1]{Section~\ref{sec:#1}}

\let\er=\eqnref
\let\el=\eqnlabel

\newcommand{\eg}{{\it e.g.\/}}
\newcommand{\ie}{{\it i.e.\/}}

\newcommand{\thmlabel}[1]{\label{thm:#1}}
\newcommand{\thmref}[1]{Theorem \ref{thm:#1}}
\newcommand{\numberedtheorem}[3]{
	\ifthen{\not\equal{#1}{}}{
		\let\oof=\thetheorem
		\renewcommand{\thetheorem}{#1}
	}
	\begin{theorem}
	\ifthen{\not\equal{#2}{}}{
		\thmlabel{#2}
	}
	#3
	\end{theorem}
	\ifthen{\not\equal{#1}{}}{
		\let\thetheorem=\oof
		\addtocounter{theorem}{-1}
	}
}
\newcommand{\lemlabel}[1]{\label{lem:#1}}
\newcommand{\lemref}[1]{Lemma \ref{lem:#1}}
\newcommand{\numberedlemma}[3]{
	\ifthen{\not\equal{#1}{}}{
		\let\oof=\thelemma
		\renewcommand{\thelemma}{#1}
	}
	\begin{lemma}
	\ifthen{\not\equal{#2}{}}{
		\lemlabel{#2}
	}
	#3
	\end{lemma}
	\ifthen{\not\equal{#1}{}}{
		\let\thelemma=\oof
		\addtocounter{lemma}{-1}
	}
}

\newif\ifjournalsub

\newcommand{\httpref}[1]{}

\newcommand{\be}[1]{\begin{equation}\eqnlabel{#1}}
\newcommand{\ee}{\end{equation}}

\newcommand{\vecphi}{\vec \phi}
\newcommand{\Dphi}{\Delta \phi}
\newcommand{\vecDphi}{\vv{\Delta \phi}}
\newcommand{\dphi}{\delta \phi}
\newcommand{\vecdphi}{\vv{\delta \phi}}
\newcommand{\ah}{\alpha_g} 
\newcommand{\tDagger}{{t^{\!\dagger}}}

\def\ssp{\def\baselinestretch{0.95645}\large\normalsize}

\ssp

\begin{document} 
\title{
Hierarchical Abstraction of Phase Response Curves of Synchronized Systems of Coupled Oscillators
}
\author{
Jaijeet Roychowdhury, {EECS Department, University of California,
Berkeley}
} 

\maketitle 
\thispagestyle{empty}

\begin{abstract}

We prove that a group of injection-locked oscillators, each
modelled using a nonlinear phase macromodel, responds as a single oscillator to
small external perturbations. More precisely, we show that any group of injection-locked oscillators has a
single effective PRC \cite{Winfree67} or PPV \cite{DeMeRoTCAS2000,DeRoTCAD2003} that characterises its 
phase/timing response to small external perturbations.
This result constitutes a foundation for
understanding and predicting synchronization/timing hierarchically in large,
complex systems that arise in nature and engineering.
  
\end{abstract}

\section{PRC/PPV Phase Macromodels} 

Given an ODE or DAE description
\be{ODE}
	\ddt{\vec q(\vec x(t))}{t} + \vec f(\vec x) + \vec b(t) = \vec 0
\ee
of an oscillator with an orbitally stable $T$-periodic autonomous solution $\vec x_s(t)$,
it can be shown \cite{DeMeRoTCAS2000,DeIJCTA2000} that the timing jitter or phase characteristics
of the oscillator, under the influence of small perturbations $\vec b(t)$, can be captured by the nonlinear scalar 
differential equation
\be{PPVeqn}
	\ddt{\alpha(t)}{t}  = \vec v_1^T(t + \alpha(t)) \cdot \vec b(t),
\ee
where the quantity $\vec v_1(\cdot)$, a $T$-periodic function of time, is known as the \textit{Phase Response Curve}
(PRC) \cite{Winfree67} or \textit{Perturbation Projection Vector} (PPV)
\cite{DeMeRoTCAS2000,DeRoTCAD2003}.

For convenience, we scale the time axis to normalize all periods to $1$.
Define a $1$-periodic version of the steady state solution to be
\be{xp}
\vec x_p(t) = \vec x_s(tT),
\ee 
and a $1$-periodic version of the PPV to be
\be{onePppv}
\vec p(t) = \vec v_1(tT).
\ee 
Using these $1$-periodic quantities and defining $f\triangleq \frac{1}{T}$, \er{PPVeqn} can be expressed as
\be{PPVeqnScaled}
	\ddt{\alpha(t)}{t}  = \vec p^T\left(ft + f\alpha(t)\right) \cdot \vec b(t).
\ee
Defining \textit{phase} to be
\be{phi}
	\phi(t) = ft + f\alpha(t),
\ee
\eqnref{PPVeqnScaled} becomes
\be{PPVeqnPhi}
	\ddt{\phi(t)}{t}  = f + f \vec p^T(\phi(t)) \cdot \vec b(t).
\ee

$x(t)$, the solution of \er{ODE}, can often be approximated usefully by 
a phase-shifted version of its unperturbed periodic solution, \ie, 
\be{xoft}
	\vec x(t) \simeq \vec x_s(T\phi(t)) = \vec x_p(\phi(t)).
\ee

\er{PPVeqn} (or equivalently, \er{PPVeqnPhi}) is termed the \textit{PPV equation} or \textit{PPV phase macromodel}.
In the absence of any perturbation $\vec b(t)$, note that
$\alpha(t) \equiv 0$ (w.l.o.g), $\vec x(t) = \vec x_s(t) = \vec x_p(ft)$ and
$\phi(t) = ft$. We will call the latter
the \textit{phase of natural oscillation} and denote it by
$\phi^\diamond(t) \triangleq ft$.

\section{Derivation of hierarchical PPV macromodel}

\subsection{Coupled Phase System and its Properties}

\subsubsection{Coupled system of PPV phase macromodels}
\begin{figure}[htbp]
\centering{
\epsfig{file=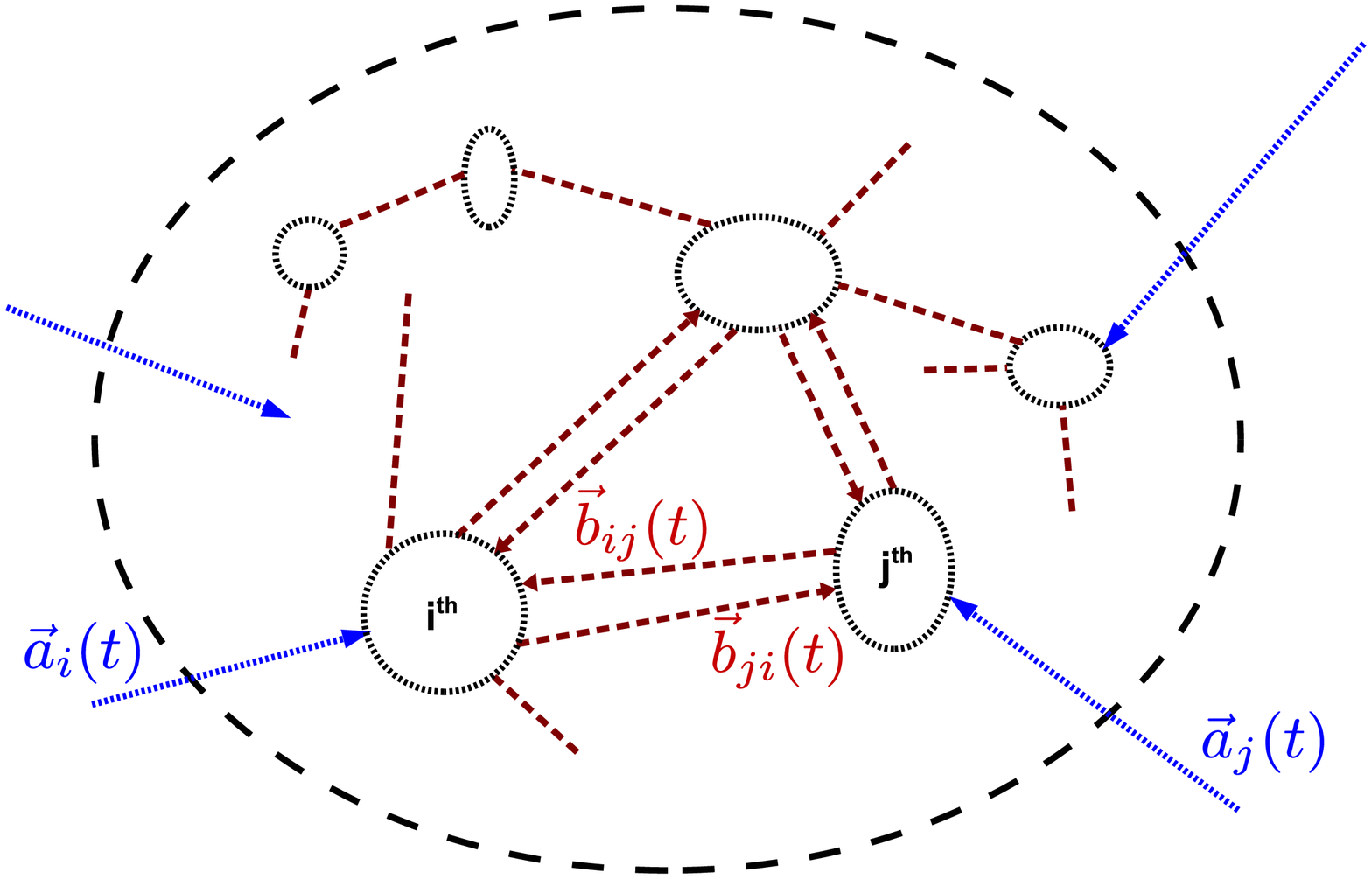,width=0.9\linewidth}
}
\caption{\figlabel{coupledoscillators}Oscillator system with internal coupling and external inputs.}
\end{figure}
Consider a group of $N \ge 2$ coupled oscillators (\figref{coupledoscillators}).
We model each oscillator by its PPV equation \eqnref{PPVeqnPhi}: 
\be{PPVeqnithphi}
	\ddt{\phi_i(t)}{t}  = f_i + f_i \vec p_i^T\left(\phi_i(t)\right) \cdot \vec b_i(t), \quad i = 1,\cdots,N,
\ee
where $i$-subscripted quantities refer to the $i^{\text{th}}$ oscillator.
Inputs to each oscillator are drawn from two sources (as depicted in \figref{coupledoscillators}): 1) internal couplings with other oscillators, and 2) external sources. 
$\vec b_i(t)$ can therefore be written as
\be{coupledOscbtphi}
	\vec b_i(t) = \vec a_i(t) + \sum_{\substack{j\not=i\\j=1}}^{N} \vec b_{ij}\left(\phi_j(t)\right),
\ee
where $\vec a_i(t)$ is the external input (\ie, from outside the group of $N$ oscillators) to the $i^{\text{th}}$ oscillator, and $\vec b_{ij}\left(\phi_j(t)\right)$
represents the influence of the $j^{\text{th}}$ oscillator on the $i^{\text{th}}$. 

We
make the natural assumption that the $\vec b_{ij}(\cdot)$ are 1-periodic --- \ie, that each oscillator generates outputs 
that follow its own phase and timing properties; it is these outputs that couple
internally to the inputs of other oscillators. Note that as $i$ varies, the dimensions of 
$\vec p_i(t)$, $\vec a_i(t)$ and $\vec b_{ij}$ can differ, since they depend on the size of 
the $i^{\text{th}}$ oscillator's differential equations.

The system of $N$ equations \eqnref{PPVeqnithphi} can be written in vector ODE form as
\be{CPS}
	\ddt{\vec \phi(t)}{t} = \vec g_{\phi}\big(\vec \phi(t)\big) + \vec b_\phi\big(\vec \phi(t), t\big),
\ee
where
\begin{align}
\el{vecphiDefn}
\vecphi(t) & \triangleq \begin{bmatrix} \phi_1(t)\\ \vdots \\ \phi_N(t)\end{bmatrix}, \\
\el{vecgphiDefn}
\vec g_\phi(\vec \phi) & \triangleq \begin{bmatrix} f_1 + f_1 \vec p_1^T(\phi_1) \cdot \left. \sum_{j=2}^{N} \vec b_{1j}\left(\phi_j(t)\right) \right.
\\ \vdots \\ f_N + f_N \vec p_N^T(\phi_N) \cdot 
\left. \sum_{j=1}^{N-1} \vec b_{Nj}\left(\phi_j(t)\right) \right. \end{bmatrix}, \\
\text{and } \vecbphi(\vecphi,t) & \triangleq \begin{bmatrix} f_1 \vec p_1^T(\phi_1) \cdot \vec a_{1}(t) \\ \vdots \\ f_N \vec p_N^T(\phi_N) \cdot \vec a_{N}(t)\end{bmatrix}.
\el{vecbphiDefn}
\end{align}
We will refer to \eqnref{CPS} as the \textit{Coupled Phase System (CPS)}.

Note that $\vec b_\phi(\vec \phi, t) \equiv \vec 0$ in the absence of inputs external to the group of oscillators, \ie, when $a_i(t) \equiv \vec 0$.
Note also that $\vec g_\phi(\vecphi)$ is $1$-periodic in each component of $\vecphi$, \ie, it is $1$-periodic in each $\phi_i$.
Such functions are termed \textit{cylindrical} \cite[page 236]{Farkas94}.

\subsubsection{Locking in the absence of external inputs}
\begin{assumption}
\asslabel{OscGroupIsInjectionLocked}
In the absence of external inputs (\ie, if $\vec a_i(t) \equiv \vec 0, \forall i$), assume that the group of $N$ oscillators locks stably\footnote{\secref{stability} will expand
on the notion of lock stability.} 
to a common frequency $f^*$ (equivalently, with a common period $T^*=\frac{1}{f^*}$). 
\end{assumption}

Denote the phase of the $i^{\text{th}}$ oscillator, when locked as in \assref{OscGroupIsInjectionLocked} to the other oscillators in the group,
by $\phi_i^*(t)$.  Note that this phase will typically be different from the oscillator's
phase of natural oscillation $\phi_i^\diamond(t) = f_i \, t$, on account of inputs via coupling from other oscillators in the group.

Denoting
\be{vecphiStarDefn}
	\vecphi^*(t) \triangleq \begin{bmatrix} \phi_1^*(t)\\ \vdots \\ \phi_N^*(t)\end{bmatrix},
\ee
note that $\vecphi^*(t)$ satisfies \eqnref{CPS} with no external inputs, \ie,
\be{CPSAutonomous}
	\ddt{\vecphi^*(t)}{t} = \vec g_{\phi}\big(\vecphi^*(t)\big).
\ee
We term $\vecphi^*(t)$ the \textit{system phase during externally-unperturbed lock}.

\subsubsection{D-periodicity of $\vecphi^*(t)$}

$T^*$-periodicity of each oscillator locked at frequency $f^*$ implies that
\be{lockedxit}
	\vec x_i(t) = \vec x_{p,i}(\phi_i^*(t))
\ee
is $T^*$-periodic $\forall i$; \ie, 
\be{xiperiodic}
	\begin{split}
	\vec x_i(t+T^*) & = \vec x_i(t), \quad \text{or}\\
	\vec x_{p,i}(\phi_i^*(t+T^*)) & = \vec x_{p,i}(\phi_i^*(t)).
	\end{split}
\ee
From definition, $\vec x_{p,i}(\cdot)$ is $1$-periodic. \eqnref{xiperiodic} is satisfied for \textit{arbitrary} 
$1$-periodic $\vec x_{p,i}(\cdot)$  iff
\be{phiTstarPeriodic}
	\phi_i^*(t+T^*) = n + \phi_i^*(t), \quad \forall t, \quad n \in \Bbb{Z}.
\ee

\par
Define the \textit{ideal phase of oscillation at frequency $f^*$}  to be
\be{phiref}
	\phistarIdeal(t) = f^* t.
\ee
$\phistarIdeal(t)$ satisfies \eqnref{phiTstarPeriodic} (with $n=1$), but note that the phase of each locked oscillator
in the system need not necessarily equal $\phistarIdeal(t)$.
A more general form for $\phi_i^*(t)$ that also satisfies \eqnref{phiTstarPeriodic} is
\be{DphiStarDefn}
	\phi_i^*(t) = \phistarIdeal(t) + \Dphi_i^*(t) = f^* t + \Dphi_i^*(t),
\ee
where $\Dphi_i^*(t)$ is itself $T^*$-periodic\footnote{$\Dphi_i^*(t)$ can in fact itself satisfy \eqnref{phiTstarPeriodic} with arbitrary $n$, but
$n\not=0$ would make the long-term frequency of $x_{p,i}(\phi_i(t))$ different from $f^*$, violating \assref{OscGroupIsInjectionLocked}.}. 
Equivalently\footnote{We use the notation, borrowed from MATLAB, that the sum of a scalar and a vector
means that the scalar is added to each element of the vector.}, 
\be{vecphiStarDperiodic}
	\vecphi^*(t) = \phistarIdeal(t) + \vecDphi^*(t) = f^*t + \vecDphi^*(t),
\ee
where
\be{vecDphiStarDefn}
	\vecDphi^*(t) \triangleq \begin{bmatrix} \Delta \phi_1^*(t)\\ \vdots \\ \Delta \phi_N^*(t)\end{bmatrix}
\ee
is $T^*$-periodic. Functions of the form \eqnref{vecphiStarDperiodic} are termed \textit{D-periodic} or \textit{derivo-periodic} with period $T^*$ \cite{Farkas94}.
$\{\Dphi_i^*(t)\}$ represent short-term phase changes within each cycle that do not affect the long-term frequency of the oscillator.

The above considerations motivate:

\begin{assumption}
\asslabel{phistarDperiodic}
$\vecphi^*(t)$, the phase of the CPS during externally-unperturbed lock, is $T^*$--D-periodic.
\end{assumption}\ \\[-2em]

\subsubsection{Arbitrary time shifts of $\vecphi^*(t)$ are also solutions of the CPS}
\begin{lemma}
\lemlabel{phiStarNotUnique}
The phase during externally-unperturbed lock, $\vecphi^*(t)$, is not unique.
Indeed, for any arbitrary time-shift $\tau$,  
\be{vecphistarTimeShifted}
	\vecphi^*(t-\tau)
\ee
solves \eqnref{CPSAutonomous}.
\begin{proof}
Follows directly from substituting \eqnref{vecphistarTimeShifted} in the autonomous system \eqnref{CPSAutonomous} and using the
facts that 1) $\vec g_\phi(\cdot)$ in \eqnref{vecgphiDefn} is cylindrical with period 1, and 2) $\vecphi^*(t)$ is $T^*$--D-periodic (\assref{phistarDperiodic}).
\end{proof}
\end{lemma}

\subsection{Periodic time-varying linearization of the CPS}
\seclabel{linearization}

\subsubsection{Linearization under small-deviation assumption}
\seclabel{CPSlinearization}
If the external inputs $\{\vec a_i(t)\}$ are small,  then $\vec b_{\phi}(\vecphi,t)$ is small and
\eqnref{CPS} constitutes a small perturbation of \eqnref{CPSAutonomous}.
We express $\vecphi(t)$, the solution of \eqnref{CPS}, as a deviation from $\vecphi^*(t)$, the solution 
of \eqnref{CPSAutonomous}:
\be{vecdphiDefn}
	\vecphi(t) = \vecphi^*(t) + \vecdphi(t).
\ee

We term $\vecdphi(t)$ the \textit{orbital deviation}. Using \eqnref{vecdphiDefn}, we now attempt to solve \eqnref{CPS} via linearization.\\

\begin{assumption}
\asslabel{vecdphiIsSmall}
$\vecdphi(t)$ remains small for all $t$, provided the external input $\vec b_{\phi}(\cdot,\cdot)$ 
is small enough for all $t$.\footnote{\ie, $\|\vecdphi(t)\| < M\|\vec b_{\phi}(\cdot,\cdot)\|$ for
some finite constant $M>0$.} 
\end{assumption}\ \\[-2em]

Applying \assref{vecdphiIsSmall}, we start the process of linearizing \eqnref{CPS}:
\be{CPSLinearizationOne}
\begin{split}
	\ddt{\vecphi^*(t)}{t} + \ddt{\vecdphi(t)}{t} & \simeq \vec g_{\phi}\big(\vecphi^*(t)\big) +
		  \frac{\partial \vec g_{\phi}}{\partial \vecphi}\big(\vecphi^*(t)\big) \vecdphi(t) \\
							& \qquad\qquad+ \vec b_{\phi}\big(\vecphi(t), t\big).
\end{split}
\ee
Using \eqnref{CPSAutonomous}, we obtain
\be{CPSLinearizationTwo}
\begin{split}
	\ddt{\vecdphi(t)}{t} & \simeq  \frac{\partial \vec g_{\phi}}{\partial \vecphi}\big(\vecphi^*(t)\big) \vecdphi(t)
					+ \vec b_{\phi}\big(\vecphi(t), t\big)\\
			     & \simeq \frac{\partial \vec g_{\phi}}{\partial \vecphi}\big(\vecphi^*(t)\big) \vecdphi(t) 
			     	+ \vec b_{\phi}\big(\vecphi^*(t), t\big) \\
				& \qquad\qquad\qquad \qquad   + \frac{\partial \vec b_{\phi}}{\partial \vecphi}\big(\vecphi^*(t), t\big) \vecdphi(t)\\
			     & = \left(\frac{\partial \vec g_{\phi}}{\partial \vecphi}\big(\vecphi^*(t)\big)
			     	 + \frac{\partial \vec b_{\phi}}{\partial \vecphi}\big(\vecphi^*(t), t\big)\right) \vecdphi(t) \\
			     	& \qquad\qquad\qquad \qquad   + \vec b_{\phi}\big(\vecphi^*(t), t\big).
\end{split}
\ee
From the definition of $\vec b_{\phi}(\cdot,\cdot)$ \eqnref{vecbphiDefn}, observe that $\frac{\partial \vec b_{\phi}}{\partial \vecphi}\big(\vecphi^*(t), t\big)$
is a diagonal matrix with entries
\[
f_i \vec{p_i'}^T\!\!(\phi_i^*(t)) \cdot \vec a_i(t),
\]
\ie, it is directly proportional to the external inputs $\{\vec a_i(t)\}$, which are small by assumption. Therefore, the product term
$\frac{\partial \vec b_{\phi}}{\partial \vecphi}\big(\vecphi^*(t), t\big)\vecdphi(t)$ in \eqnref{CPSLinearizationTwo} is of second order and
can be dropped from the linearization. Applying this observation and denoting
\begin{align}
\el{JstarphiDefn}
	\Jstarphi(t) & \triangleq \frac{\partial \vec g_{\phi}}{\partial \vecphi}\big(\vecphi^*(t), t\big), \text{ and}\\
	\vecbext(t) & \triangleq \vec b_{\phi}\big(\vecphi^*(t), t\big),
\end{align}
\eqnref{CPSLinearizationTwo} becomes
\be{linearizedCPS}
\boxed{
	\ddt{\vecdphi(t)}{t} \simeq \Jstarphi(t) \vecdphi(t) + \vecbext(t)
}\,.
\ee
\eqnref{linearizedCPS} is the linearization of the CPS \eqnref{CPS} around its externally-unperturbed solution $\vecphi^*(t)$.

\subsubsection{$T^*$-periodicity of $\Jstarphi(t)$}
From \eqnref{vecgphiDefn}, we can obtain expressions for the entries of $\Jstarphi(t)$. The diagonal entries of $\Jstarphi$ are
\be{JstarDphiEntriesDiagonal}
{\Jstarphi}_{\,i,i}(t) = f_i \vec{p_i'}^T\!\!(\phi_i^*(t)) \cdot \left. \sum_{\substack{j\not=i\\j=1}}^{N} \vec b_{ij}\left(\phi_j^*(t)\right) \right.,
\ee
while the off-diagonal entries are
\be{JstarDphiEntriesOffDiagonal}
{\Jstarphi}_{\,i,j}(t) = f_i \vec{p_i}^T\!\!(\phi_i^*(t)) \cdot \left. \vec{b'_{ij}}\left(\phi_j^*(t)\right) \right..
\ee
Because of the $1$-periodicity of $\vec p_i(\cdot)$ and $\vec b_{ij}(\cdot)$, and the $T^*$--D-periodicity of $\phi_i^*(t)$,
each entry of $\Jstarphi$ is $T^*$-periodic, hence the entire matrix function $\Jstarphi(t)$ is $T^*$-periodic. The linearized CPS \eqnref{linearizedCPS} is
therefore periodically time varying, \ie, it is a \textit{linear periodically time varying (LPTV)} system.

\subsection{$T^*$-periodic homogeneous solution of the linearized CPS}
\begin{lemma}
\lemlabel{linearizedCPShomogeneousSolution}
The homogeneous part of the linearized CPS \eqnref{linearizedCPS}, \ie, 
\be{linearizedCPShomogenous}
	\ddt{\vecdphi(t)}{t} = \Jstarphi(t) \,\, \vecdphi(t),
\ee
has the $T^*$-periodic solution
\be{vecdphiStarDefn}
	\vecdphi^*(t) \triangleq \ddt{\vecphi^*(t)}{t}.
\ee
\begin{proof}
Follows immediately from differentiating \eqnref{CPSAutonomous}.
Note that $T^*$--D-periodicity of $\vecphi^*(t)$ immediately implies that $\vecdphi^*(t)$ in \eqnref{vecdphiStarDefn} is $T^*$-periodic, since
\be{vecdphiStarExpanded}
	\vecdphi^*(t) = f^* + \ddt{\vecDphi^*(t)}{t},
\ee
with the latter term $T^*$-periodic.
\end{proof}
\end{lemma}

\subsection{Floquet-theoretic solution of the linearized CPS}
\seclabel{Floquet}
Floquet theory \cite{Farkas94} provides an analytical form\footnote{\eqnref{FloquetSolution} holds for the case where the Floquet multipliers \eqnref{DoftDefn} are distinct,
but subject to \assref{OnlyOneFMisOne}, all subsequent results in this section remain valid for the case of repeated Floquet multipliers.} 
for the solution of \eqnref{linearizedCPS}:
\be{FloquetSolution}
	\begin{split}
	\vv{\dphi}(t) & = U(t) D(t-t_0) V^T(t_0) \vv{\dphi}_0  +\\ 
			& \qquad \qquad U(t) \int_{t_0}^t D(t-\tau) V^T(\tau) \, \vecbext(\tau) \, d\tau.
	\end{split}
\ee
$U(t)$ and $V^T(t)$ are $T^*$-periodic matrix functions, of size $N\times N$, that satisfy
\be{UVTeqI}
	U(t) V^T(t) = V^T(t) U(t) = I_{N\times N}.
\ee
\eqnref{UVTeqI} implies that the columns of $U$ and $V$ are bi-orthogonal, \ie,
\be{UVcolsBiorthogonal}
	\vec v_i^T(t) \cdot \vec u_j(t) = \delta_{ij}, \quad i,j=1,\cdots,N.
\ee
\eqnref{UVTeqI} can be written more explicitly, showing $\vec v_i$ and $\vec u_j$, as
\be{UVTeqIexpanded}
	\begin{split}
	\begin{pmatrix}
	\textcolor{blue}{\cdots\vec v_1^T(t)  \cdots} \\
	  \vdots  \\
	 \cdots\vec v_N^T(t)  \cdots 
	\end{pmatrix}
	\begin{pmatrix}
	\textcolor{blue}{\vdots} & & \vdots \\
	\textcolor{blue}{\vec u_1(t)} & \cdots & \vec u_N(t) \\
	\textcolor{blue}{\vdots} & & \vdots
	\end{pmatrix}
	\equiv
	\begin{pmatrix}
	\textcolor{blue}{1} & \\
	    & \ddots & \\
	    &             & 1
	\end{pmatrix}.
	\end{split}
\ee
Note that, in particular,
\be{v1Tdotu1Eq1}
	\vec v_1^T(t) \cdot \vec u_1(t) \equiv 1, \forall t,
\ee
a relation we will rely on later.

$D(\cdot)$ in \eqnref{FloquetSolution} is a \textit{diagonal} matrix of the form
\be{DoftDefn}
	D(t)= \begin{pmatrix} \textcolor{blue}{e^{\mu_1 t}} & & \\ & \ddots  &  \\ & & e^{\mu_N t} \end{pmatrix},
\ee
where $\{\mu_i\}$ are termed \textit{Floquet} or \textit{characteristic exponents}, and 
\be{rhoiDefn}
	\rho_i \triangleq e^{\mu_i T^*}, \quad i=1,\cdots,N,
\ee
are known as \textit{Floquet} or \textit{characteristic multipliers}.  Note that $D(t)$ is not periodic.

\eqnref{FloquetSolution} can be rewritten using $\vec v_i(\cdot)$ and $\vec u_i(\cdot)$ as
\be{FloquetSolutionExpanded}
	\begin{split}
	\vv{\dphi}(t) & = \sum_{i=1}^N \vec u_i(t) \underbrace{e^{\mu_i (t-t_0)} \vec v_i^T(t_0) \cdot \vv{\dphi}_0}_{\text{scalar}} +\\
		      & \qquad\qquad \sum_{i=1}^N \vec u_i(t) \underbrace{\int_{t_0}^t e^{\mu_i (t-\tau)} \vec v_i^T(\tau) \cdot \vecbext(\tau) \,\, d\tau}_{\text{scalar}}.
	\end{split}
\ee

\subsection{Conditions on Floquet multipliers; stability and isolation of $\vecphi^*(t)$}
\seclabel{stability}

\lemref{phiStarNotUnique} suggests that the stability of the CPS \eqnref{CPS} around its locked solution \eqnref{vecphiStarDefn} in the absence of external inputs 
is of an orbital nature \cite[Definition 5.1.1]{Farkas94} and not, \eg, of a Lyapunov one \cite[Definition
1.4.1]{Farkas94}. 

\begin{lemma}
\lemlabel{AtLeastOneFMisOne}
At least one of the Floquet multipliers $\{\rho_i\}$ \er{rhoiDefn} is 1 (equivalently, at least one of the Floquet 
exponent $\{\mu_i\}$ is 0).
\begin{proof}
Follows from the existence of the $T^*$-periodic homogeneous solution of the linearized CPS
\er{vecdphiStarExpanded}, established in \lemref{linearizedCPShomogeneousSolution}, and \cite[Corollary
2.2.3]{Farkas94}.
\end{proof}
\end{lemma}

\begin{lemma}
\lemlabel{NoFMisGreaterThanOne}
$|\rho_i| \le 1, \forall i$ (equivalently, $\Real(\mu_i) \le 0$).
\begin{proof}
Follows from \assref{OscGroupIsInjectionLocked} (\ie, that the externally-unperturbed oscillator system is mutually
injection locked in a \textit{stable} orbit) and \cite[Theorem 5.1.3]{Farkas94}.
\end{proof}
\end{lemma}

We now make an additional technical assumption regarding the Floquet multipliers:
\begin{assumption}
\asslabel{OnlyOneFMisOne}
\textit{Only} one Floquet multiplier ($\rho_1$, without loss of generality) is 1 (equivalently, w.l.o.g, $\mu_1=0$).
\end{assumption}

Note that \assref{OnlyOneFMisOne} strengthens \lemref{AtLeastOneFMisOne}. There are several factors that 
motivate this assumption:
\begin{enumerate}
	\item \assref{OnlyOneFMisOne}, together with \lemref{NoFMisGreaterThanOne}, constitute \textit{sufficient} 
		conditions for the Andronov-Witt theorem \cite[Theorems 5.3.8 and 5.1.2]{Farkas94} to hold. 
		The Andronov-Witt theorem guarantees that the CPS \eqnref{CPS}
	 	is \textit{asymptotically orbitally stable (a.o.s)} \cite[Definition 5.1.2]{Farkas94} with 
		the \textit{asymptotic phase property (a.o.p)} \cite[Definition 5.1.3]{Farkas94}.
		These properties are central to the intuitive concept of stable lock, assumed 
		in \assref{OscGroupIsInjectionLocked} and typically valid in most applications.
	\item \assref{OnlyOneFMisOne} also constitutes a sufficient condition for the orbit $\vecphi^*(t)$
		\er{vecphiStarDefn} to be \textit{isolated}, \ie, not embedded in a continuum of orbits with
		continuously-varying periods \cite[Theorems 5.3.9 and 5.2.3]{Farkas94}. Isolation is also central to
		the intuitive notion of stable lock.
	\item Although \assref{OnlyOneFMisOne} is not a \textit{necessary} condition for asymptotic 
		orbital stability with the asymptotic phase property, or for isolation, oscillators that do not
		satisfy the assumption while still being a.o.s + a.o.p tend to be ``unnatural''.  For example,
		\cite[Example 5.2.2]{Farkas94} features an orbit that is a.o.s + a.o.p but has three repeated Floquet
		multipliers that equal $1$; however, this orbit is not periodic.
\end{enumerate}
Thus, in most practical situations, \assref{OnlyOneFMisOne} is essentially equivalent to the oscillator group's being
``stably locked''.

\subsection{\assref{vecdphiIsSmall} (deviations are small) is invalid}
\seclabel{linearizationInvalid}

\begin{lemma}
\lemlabel{smallDeviationAssumptionInvalid}
\assref{vecdphiIsSmall} is invalid.
\begin{IEEEproof}
Using \assref{OnlyOneFMisOne}, the second summation term of \er{FloquetSolutionExpanded}, which
captures the linearized system's response to external perturbations $\vecbext(t)$, can be written as
\be{FloquetSolutionParticularPart}
	\begin{split}
	\vec u_1(t) & \overbrace{\int_{t_0}^t  \vec v_1^T(\tau) \cdot \vecbext(\tau) \,\, d\tau}^{\text{scalar }c_1(t)}
	 \\
	& \qquad \qquad +\sum_{i=2}^N \vec u_i(t) \underbrace{\int_{t_0}^t e^{\mu_i (t-\tau)} \vec v_i^T(\tau) \cdot \vecbext(\tau) \,\, d\tau}_{\text{scalar } c_i(t)}.
	\end{split}
\ee
The first term is of the form $c_1(t) \vec u_1(t)$, where $c_1(t)$ is the scalar
\be{FloquetSolutionBlowup}
c_1(t) = \int_{t_0}^t  \vec v_1^T(\tau) \cdot \vecbext(\tau) \,\, d\tau.
\ee
Because $\vec v_1(\tau)$ is periodic, there exist many possibilities for small $\vecbext(t)$ that make $c_1(t)$ increase indefinitely and without bound as $t$ increases. 
For example, if $\vecbext(t) = \epsilon \vec u_1(t)$, with $\epsilon\not=0$
being any constant, then, from \eqnref{v1Tdotu1Eq1}, $c_1(t) = (t-t_0)\epsilon$; 
 \ie, $c_1(t)$ increases without bound.
The remaining terms in \er{FloquetSolutionParticularPart} are bounded because $\Real(\mu_i) < 0, \quad \forall i>1$, hence cannot cancel the unbounded
increase of the first term.

In other words, \assref{vecdphiIsSmall}, upon which the linearized system \eqnref{linearizedCPS}, its solution  \er{FloquetSolution}, and indeed, the expression for the unbounded term $c_1(t)$
in \eqnref{FloquetSolutionBlowup} all depend, is violated. Thus we have arrived at a contradiction, implying that the original premise
\assref{vecdphiIsSmall} must be invalid (subject to the other assumptions' validity). 
\end{IEEEproof}
\end{lemma}
That deviations can grow to be large even when external inputs remain small is a manifestation of the inherently marginal nature of orbital
stability, \ie, that $\mu_1=0$.

\subsection{Time-shifted perturbed response assumption}

\begin{lemma}
\lemlabel{u1EQvecdphiStar}
Without loss of generality,
\be{u1EQvecdphiStar}
	\vec u_1(t) = \vv{\dphi}^*(t) = \ddt{\vec \phi^*(t)}{t}.
\ee
\begin{IEEEproof}
The first summation term in \eqnref{FloquetSolutionExpanded}, \ie,
\be{FloquetSolutionHomogeneousPart}
\sum_{i=1}^N \vec u_i(t) \underbrace{e^{\mu_i (t-t_0)} \vec v_i^T(t_0) \cdot \vv{\dphi}_0}_{\text{scalar}},
\ee
represents a general solution of \eqnref{linearizedCPShomogenous}.
We already know that $\vv{\dphi}^*(t)$ \eqnref{vecdphiStarDefn} is a nontrivial \textit{periodic} solution of \eqnref{linearizedCPShomogenous}.
Using $\rho_1=1$ from \assref{OnlyOneFMisOne}, this periodic solution must equal the $i=1$ term in \eqnref{FloquetSolutionHomogeneousPart}, since (also from
\assref{OnlyOneFMisOne}) the remaining terms for $i=2,\cdots,N$ are not periodic and indeed, decay to $0$ as $t\to \infty$. Hence we have
\be{u1EQvecdphiStarScaled}
\vv{\dphi}^*(t)=k_2\vec u_1(t) \underbrace{\vec v_1^T(t_0) \cdot \vv{\dphi}^*(t_0)}_{\text{scalar constant } k_1}.
\ee
where $k_2$ is an arbitrary scalar constant. Note that $k_1 \not=0$, otherwise \eqnref{FloquetSolutionHomogeneousPart} would
be identically zero, hence would not match any nontrivial $\vv{\dphi}^*(t)$. Choosing $k_2 = \frac{1}{k_1}$ (without loss
of generality, since $\vec v_1(t)$ can be scaled to satisfy \eqnref{v1Tdotu1Eq1}) results in \eqnref{u1EQvecdphiStar}.
\end{IEEEproof}
\end{lemma}
Geometrically, $\frac{d}{dt} \vecphi^*(t)$ is the tangent to the externally-unperturbed orbit of the locked system in phase space; \lemref{u1EQvecdphiStar} thus justifies the terminology \textit{tangent
vector} for $\vec u_1(t)$.

Attempting to restore validity to the failed linearization procedure above,
observe that using \eqnref{u1EQvecdphiStar}, the unbounded term in \eqnref{FloquetSolutionBlowup} can be written as 
\be{UnboundedTermWithXsdot}
	c_1(t) \vec u_1(t) = c_1(t) \ddt{\vec \phi^*(t)}{t}.
\ee
Observe also that if $c_1(t)$ \textit{were} bounded and small, then 
\be{timeShiftMotivation}
	 \vec \phi^*(t) + c_1(t) \ddt{\vec \phi^*(t)}{t} \simeq \vec \phi^*\big(t + c_1(t)\big),
\ee
to first order.
This suggests that the unboundedly growing component of $\vv{\dphi}(t)$ in \er{FloquetSolutionExpanded} may be the manifestation
of a time-shift to the unperturbed solution $\vec \phi^*(t)$. A time shift along the orbit is also
suggested by the definition of orbital stability \cite[Definition 5.1.1]{Farkas94} and by the physical intuition that autonomous oscillators,
having no intrinsic ``time reference'', can 
``slip in phase'', \ie, they cannot correct errors in phase. Accordingly, we modify the assumed form of 
the perturbed solution \eqnref{vecdphiDefn} to
\begin{assumption}
\asslabel{TimeShiftedDeviationForm}
\be{TimeShiftedDeviationForm}
	\vec \phi(t) = \vecphi^*\big(t+\ah(t)\big) + {\vecdphi}(t),
\ee
where ${\vecdphi}(t)$ remains small for all time (\ie, $\|{\vecdphi}(t)\| < M \| \vecbext(t)\|$ for some finite $M>0$).
\end{assumption}\ \\[-1.0em]
$\ah(t)$ is a (yet-to-be-determined) time shift that can depend on the input perturbation $\vecbext(t)$ and can grow unboundedly with time.
Importantly, we have retained \assref{vecdphiIsSmall}, \ie, that $\vecdphi(t)$ in \er{TimeShiftedDeviationForm}
remains bounded and small for all time.

We shall prove that unlike \eqnref{vecdphiDefn}, the time-shifted deviation form \eqnref{TimeShiftedDeviationForm} will allow $\vecdphi(t)$ to remain bounded and
small, providing the time-shift $\ah(t)$ is chosen appropriately.

\subsection{Base for time-shifted linearization}
In \secref{linearization}, the CPS was linearized around the unperturbed orbit $\vecphi^*(t)$. 
The process of linearization relied on the fact that $\vecphi^*(t)$
satisfied \er{CPSAutonomous}. We would like to find a replacement for \eqnref{CPSAutonomous}
that is satisfied by 
\be{TimeShiftedVecPhiStar}
	\vecphi(t) = \vecphi^*(t+\ah(t))
\ee
instead.

\begin{lemma}
\lemlabel{TSlinearizationBaseCase}
Given any scalar, differentiable function $\alpha_g(t)$, the CPS \er{CPS} is solved exactly by
\er{TimeShiftedVecPhiStar} for perturbations of the form
\be{bAlongU1t}
\vecbphi(\vecphi(t),t) \triangleq K(t) \vec u_1(t+\ah(t)),
\ee
where $K(t) \equiv \frac{d}{dt}\ah(t)$.
\begin{proof}
Denoting ``shifted time'' to be
\be{tDaggerDefn}
\tDagger \triangleq t + \ah(t),
\ee
substituting \er{TimeShiftedVecPhiStar} and \er{bAlongU1t} into \er{CPS} and simplifying using \eqnref{CPSAutonomous} and
\eqnref{u1EQvecdphiStar}, we obtain
\be{b1AlphaHrelationOne}
	\begin{split}
	&(1+\dot\ah(t))\ddt{\vec \phi^*\big(\tDagger\big)}{\tDagger}  = \vec g_{\phi}\big(\vec \phi^*(\tDagger)\big) + K(t) \vec u_1(\tDagger)\\
	\Rightarrow &\dot\ah(t)\ddt{\vec \phi^*\big(\tDagger\big)}{\tDagger}  = K(t) \vec u_1(\tDagger) \\
	\Rightarrow &\dot\ah(t) \vec u_1(\tDagger)  = K(t) \vec u_1(\tDagger)\\
	\Rightarrow &\dot\ah(t) \vec u_1(t+\ah(t))  = K(t) \vec u_1(t+\ah(t)).
	\end{split}
\ee
\er{b1AlphaHrelationOne} is always satisfied if $\ah(t)$ and $K(t)$ are related by
\be{b1AlphaHrelation}
	\dot\ah(t) = K(t).
\ee
\end{proof}
\end{lemma}

\subsection{Time-shifted linearization}
\seclabel{TSlinearization}
We proceed to linearize the CPS \er{CPS} around solutions of the form \er{TimeShiftedVecPhiStar}.
To this end, we split the external input $\vecbphi(\vecphi,t)$ \er{vecbphiDefn} into two parts:
\be{vecbphiSplit}
	\vecbphi\big(\vecphi,t\big) = {\vecbphi}_{\!1}(t)+ {\vecbphi}_{2}\big(\vecphi,t\big),
\ee
with the intent that if only the first component ${\vecbphi}_1(t)$ is retained, then
\er{TimeShiftedVecPhiStar} should solve the CPS \er{CPS} \textit{exactly}, \ie,
\be{CPSwithVecbphi1Only}
	\ddt{\vecphi^*\big(t+\ah(t)\big)}{t} = \vec g_{\phi}\big(\vecphi^*(t+\ah(t))\big) + {\vecbphi}_{\!1}(t).
\ee

Motivated by \lemref{TSlinearizationBaseCase},
we explore perturbations of the form \er{bAlongU1t} along the tangent vector, \ie, of the form
\be{b1oftForm}
	{\vecbphi}_1(t) = K(t) \vec u_1\big(t+\ah(t)\big).
\ee
Given any small external perturbation $\vecbphi\big(\vecphi(t),t\big)$ \er{vecbphiDefn}, our
goal is to find such an $\ah(t)$ (and, using \lemref{TSlinearizationBaseCase}, its derivative $K(t)$) that ${\vecbphi}_2(\cdot,\cdot)$ in \er{vecbphiSplit}, as well as the orbital deviation $\vecdphi(t)$
in \er{TimeShiftedDeviationForm}, both remain small.

The flow of the time-shifted linearization procedure is:
\begin{enumerate}
	\item Start by assuming \textit{any} scalar function $K(t)$;
	\item Define $\ah(t)$ using \er{b1AlphaHrelation}, \ie, $\ddt{\ah(t)}{t} = K(t)$;
	\item Define ${\vecbphi}_{\!1}(t)$ using \er{b1oftForm};
	\item Incorporate the split-up form \er{vecbphiSplit} of the external input perturbation in the CPS \er{CPS}; \label{step4}
	\item Assuming a solution of the form \er{TimeShiftedDeviationForm} in \assref{TimeShiftedDeviationForm}, 
		linearize \er{CPS} using \er{CPSwithVecbphi1Only} as the base case; and
	\item Using the solution of the above linearization, obtain a constraint on $\ah(t)$ 
		(equivalently, on $K(t)$) under which \assref{TimeShiftedDeviationForm} holds with 
		$\vecdphi(t)$ bounded and small for all time. The equation specifying this constraint will turn out
		to be of central importance, in that it governs the phase/timing responses of the injection-locked
		system of oscillators to external perturbations.
	\item When $\ah(t)$ (equivalently, its derivative $K(t)$) is chosen to satisfy the above constraint, show
		that the phase deviation $\vecdphi(t)$ in \eqnref{TimeShiftedDeviationForm} always remains small, thus
		validating \assref{TimeShiftedDeviationForm} and the entire time-shifted linearization procedure.
\end{enumerate}

Starting from Step \ref{step4}, write the CPS \er{CPS} as
\be{CPSinputSplit}
	\ddt{\vecphi(t)}{t} = \vec g_{\phi}\big(\vecphi(t)\big) + {\vecbphi}_{\!1}(t)+ {\vecbphi}_{\!2}\big(\vecphi(t),t\big).
\ee

Incorporating \er{TimeShiftedDeviationForm}, \er{b1AlphaHrelation} and \er{b1oftForm} in \eqnref{CPSinputSplit}, we obtain
\be{CPStsLinearization1}
	\begin{split}
	& \ddt{\big[\vecphi^*(t+\ah(t)) + \vecdphi(t)\big]}{t} = \vec g_{\phi}\big(\vecphi^*(t+\ah(t)) + \vecdphi(t)\big) \\
		&  \qquad\qquad
		+ K(t) \vec u_1(t+\ah(t))
		+ {\vecbphi}_{\!2}\big(\vecphi^*(t+\ah(t)) + \vecdphi(t),t\big).
	\end{split}
\ee

Linearizing $\vec g_{\phi}(\cdot)$ in \eqnref{CPStsLinearization1}, we obtain
\be{CPStsLinearization2}
	\begin{split}
	& \ddt{\vecphi^*(t+\ah(t))}{t} + \ddt{\vecdphi(t)}{t} = \vec g_{\phi}\big(\vecphi^*(t+\ah(t))\big)  \\
		& \qquad\qquad+ \Jstarphi\big(t+\ah(t)\big)\vecdphi(t) 
		+ K(t) \vec u_1(t+\ah(t)) \\
		& \qquad\qquad\qquad\qquad\qquad+ {\vecbphi}_{\!2}\big(\vecphi^*(t+\ah(t)) + \vecdphi(t),t\big).
	\end{split}
\ee

Applying the base for time-shifted linearization \er{CPSwithVecbphi1Only} and our proposed form \er{b1oftForm} for ${\vecbphi}_{\!1}(t)$, \er{CPStsLinearization2} can be
simplified to
\be{CPStsLinearization3}
	\begin{split}
	 \hskip-0.7em\ddt{\vecdphi(t)}{t} =   \Jstarphi\big(t+\ah(t)\big)\vecdphi(t) 
		  + {\vecbphi}_{\!2}\big(\vecphi^*(t+\ah(t)) + \vecdphi(t),t\big).
	\end{split}
\ee

Observe that from definition \er{vecbphiSplit}, \er{b1oftForm},
\be{CPStsLinearization4}
	\begin{split}
	{\vecbphi}_{\!2}\big(\vecphi^*(t+\ah(t)) + \vecdphi(t),t\big)  = &\vecbphi(\vecphi^*(t+\ah(t)) + \vecdphi(t),t)\\
		&\qquad - K(t) \vec u_1(t+\ah(t)),
	\end{split}
\ee
hence \er{CPStsLinearization3} can be written as
\be{CPStsLinearization5}
	\begin{split}
	\ddt{\vecdphi(t)}{t} & =   \Jstarphi\big(t+\ah(t)\big)\vecdphi(t) - K(t) \vec u_1(t+\ah(t)) \\
	& \qquad + \vecbphi(\vecphi^*(t+\ah(t)) + \vecdphi(t),t).
	\end{split}
\ee
Using the same reasoning as for \er{CPSLinearizationTwo} in \secref{CPSlinearization}, $\vecdphi(t)$ in
the last term of \er{CPStsLinearization5} can be dropped because it contributes only a second-order term to the
linearization. Hence \er{CPStsLinearization5} becomes
\be{tsLinearizedCPS}
	\begin{split}
	\ddt{\vecdphi(t)}{t} & =   \Jstarphi\big(t+\ah(t)\big)\vecdphi(t) - K(t) \vec u_1(t+\ah(t)) \\ & \qquad + \vecbphi(\vecphi^*(t+\ah(t)),t) \\
			     & =  \Jstarphi\big(\tDagger\big)\vecdphi(t) + \vecbphi(\vecphi^*(\tDagger),t)- K(t) \vec u_1(\tDagger) \\
			     & =  \Jstarphi\big(\tDagger\big)\vecdphi(t) + {\vecbphi}_{\!2}(\vecphi^*(\tDagger),t),
	\end{split}
\ee
where we have used the notation $\tDagger$, defined in \er{tDaggerDefn}, for shifted time.

\subsection{Recasting time-shifted linearization in LPTV form}

We would now like to obtain an analytical solution of \er{tsLinearizedCPS} and use it to validate that $\vecdphi(t)$ remains small for all time. However, two
differences between \er{linearizedCPS} and \er{tsLinearizedCPS} make this more involved than for \er{linearizedCPS} in \secref{Floquet}:
\begin{enumerate}
\item The input to \er{tsLinearizedCPS} is ${\vecbphi}_{\!2}(\cdot,\cdot)$, not ${\vecbphi}(\cdot,\cdot)$ as in \er{linearizedCPS}. Whereas the latter is known
	small (due to the assumption of small external perturbations $\{a_i(t)\}$ in \er{vecbphiDefn}), there is no guarantee that ${\vecbphi}_{\!2}(\cdot,\cdot)$
	is also small. Ensuring that ${\vecbphi}_{\!2}(\cdot,\cdot)$ is small is important: if even the input to \er{tsLinearizedCPS} cannot be guaranteed small, it 
	is unreasonable to expect that its solution will remain small for all time. 
\item Unlike \er{linearizedCPS}, which is LPTV, \er{tsLinearizedCPS} is \textit{not} LPTV because though $\Jstarphi(t)$ is $T^*$-periodic, $\Jstarphi\big(t+\ah(t)\big)$ is
	not, except for special choices such as $\ah(t) \equiv 0$. We are interested in a solution of \er{tsLinearizedCPS} that is valid for any
	$\ah(t)$ (equivalently, any $K(t)$), if possible. Because \er{tsLinearizedCPS} is not LPTV, the Floquet expressions in \secref{Floquet} do not apply
	directly.
\end{enumerate}

Both issues can be addressed by restricting $\dot\ah(t) \equiv K(t)$ to be small. We state this as an assumption for the
moment\footnote{We will establish later that this assumption is in fact a consequence of the external inputs ${\vecbphi}(\cdot,\cdot)$ being small.}:
\begin{assumption}
	\asslabel{KtIsSmall}
\be{Ktdefn}
	K(t) \triangleq \dot\ah(t)
\ee
	is small and bounded with respect to $\vecbphi(\cdot,\cdot)$ \eqnref{vecbphiDefn} 
		for all time. In particular, $|K(t)| \ll 1$.
\end{assumption}

The first consequence of \assref{KtIsSmall} is that it becomes possible to guarantee that ${\vecbphi}_{\!2}(\cdot,\cdot)$, the input to the time-shifted linearization
\er{tsLinearizedCPS}, is small:
\begin{lemma}
\lemlabel{vecbhi2isSmall}
${\vecbphi}_{\!2}(\vecphi^*(\tDagger),t)$ is small for all $\tDagger, t$.
\begin{proof}
From definition \er{vecbphiSplit}, 
\be{vecbphiTwoDefn}
{\vecbphi}_{\!2}(\vecphi^*(\tDagger),t) = \vecbphi(\phi^*(\tDagger),t) - K(t) \vec u_1(\tDagger).
\ee
The first term is small from our underlying assumption of small external perturbations. The tangent vector $\vec u_1(t)$ is a periodic, bounded quantity, hence under
\assref{KtIsSmall}, the second term is also small.
\end{proof}
\end{lemma}

Another important consequence of \assref{KtIsSmall} is
\begin{lemma}
\lemlabel{ttotDaggerIsInvertible}
The mapping \er{tDaggerDefn} 
\[
t \mapsto \tDagger, \text{ \ie, } \tDagger(t) \triangleq t + \ah(t)
\]
is invertible.
\begin{proof}
It suffices to show that the mapping is a monotonically increasing one, \ie, its derivative is always positive. We have
\[
\ddt{\tDagger(t)}{t} = 1 + \dot\ah(t) = 1 + K(t).
\]
From \assref{KtIsSmall}, $|K(t)| < 1$, hence $\ddt{\tDagger(t)}{t} > 0$, \ie, $\tDagger(t)$ is monotonically increasing.
\end{proof}
\end{lemma}

We now make the following definitions:
\begin{align}
	\vecdphi^\dagger (\tDagger) & \triangleq \vecdphi(t), \eqnlabel{vecdphidaggerDefn}\\
	{\vecbphi^\dagger}_{\!2}(a,\tDagger) & \triangleq {\vecbphi}_{\!2}(a,t),  
		\text{ and}\eqnlabel{bphitwodaggerDefn}\\
	\vecbphi^\dagger(a,\tDagger) & \triangleq \vecbphi(a,t)\eqnlabel{bphidaggerDefn}.
\end{align}
The significance of the invertibility of shifted time $\tDagger$, as established by \lemref{ttotDaggerIsInvertible}, lies
in that the above definitions become possible: given any $\tDagger$, a unique $t$ is available for use in the
right hand sides of the above definitions.

Using the above definitions, \er{tsLinearizedCPS} can be expressed using $\tDagger$ as
\be{LPTVtsLinearizedCPSminusOne}
	 \left(1+\dot\ah(t)\right)\ddt{\vecdphi^\dagger(\tDagger)}{\tDagger} =   \Jstarphi(\tDagger)\,\vecdphi^\dagger\!(\tDagger)
		 + {\vecbphi^\dagger}_{\!2}(\vecphi^*(\tDagger),\tDagger).
\ee

We now make a technical assumption for the moment, the validity of which will be demonstrated later:
\begin{assumption}
	\asslabel{dvecdphiIsSmall}
	$\|\ddt{\vecdphi(t)}{t}\| < M \|\vecdphi(t)\|$ for some $0 < M < \infty$.
\end{assumption}

\assref{dvecdphiIsSmall} implies that the magnitude of $\ddt{\vecdphi(t)}{t}$ is within a constant factor of the magnitude of $\vecdphi(t)$, \ie, the two are of
the same order of magnitude. Intuitively, it implies that $\vecdphi(t)$ has a bounded rate of change. It follows that $\vecdphi^\dagger(\tDagger)$ also has
a bounded rate of change:
\begin{lemma}
	\lemlabel{dvecdphitdaggerIsSmall}
	$\|\ddt{\vecdphi^\dagger(\tDagger)}{\tDagger}\| < M_2 \|\vecdphi^\dagger(\tDagger)\|$ for some $0 < M_2 < \infty$.
\begin{proof}
Using \er{vecdphidaggerDefn}, \er{tDaggerDefn} and \er{b1AlphaHrelation}, we have
\[
\ddt{\vecdphi^\dagger(\tDagger)}{\tDagger} = \ddt{\vecdphi(t)}{\tDagger} = \frac{1}{1+K(t)}\ddt{\vecdphi(t)}{t}.
\]
Since $|K(t)| \ll 1$ from \assref{KtIsSmall}, we have $\|\ddt{\vecdphi^\dagger(\tDagger)}{\tDagger}\| < m \|\ddt{\vecdphi(t)}{t}\|$ for some $m < \infty$.
Using \assref{dvecdphiIsSmall}, the result follows.
\end{proof}
\end{lemma}

From \lemref{dvecdphitdaggerIsSmall}, it is apparent that the term $\dot \alpha_g \ddt{\vecdphi^\dagger(\tDagger)}{\tDagger}$ in \er{LPTVtsLinearizedCPSminusOne}
is of second order, hence can be dropped from the linearization. As a result, \er{LPTVtsLinearizedCPSminusOne} becomes

\be{LPTVtsLinearizedCPS}
	 \ddt{\vecdphi^\dagger(\tDagger)}{\tDagger} =   \Jstarphi(\tDagger)\,\vecdphi^\dagger\!(\tDagger) + {\vecbphi^\dagger}_{\!2}(\vecphi^*(\tDagger),\tDagger).
\ee
Note that \er{LPTVtsLinearizedCPS} is an LPTV system with period $T^*$.

\subsection{Floquet solution of time-shifted LPTV system}
Since \er{LPTVtsLinearizedCPS} is LPTV, the Floquet expressions in \secref{Floquet} apply, with $t$ and $\vecdphi(t)$
replaced by $\tDagger$ and $\vecdphi^\dagger(\tDagger)$,
respectively, and $\vecbext(t)$ replaced by ${\vecbphi^\dagger}_{\!2}(\vecphi^*(\tDagger),\tDagger)$.
\er{FloquetSolutionExpanded} becomes
\be{FloquetSolutionOfLPTVtsLinearizedCPS}
	\begin{split}
	\vecdphi^\dagger(\tDagger) & = \sum_{i=1}^N \vec u_i(\tDagger) \underbrace{e^{\mu_i (\tDagger-t^\dagger_0)} \vec v_i^T(t^\dagger_0) \cdot \vecdphi^\dagger(t_0^\dagger)}_{\text{scalar}} +\\
		      & \qquad \sum_{i=1}^N \vec u_i(\tDagger) \underbrace{\int_{t_0^\dagger}^\tDagger e^{\mu_i (\tDagger-\tau)} \vec v_i^T(\tau) \cdot 
		      	{\vecbphi^\dagger}_{\!2}(\vecphi^*(\tau),\tau)\,\, d\tau}_{\text{scalar}},
	\end{split}
\ee
while the term $c_1(t)$ in \er{FloquetSolutionBlowup}, which causes unbounded growth and resulting breakdown
of linearization, becomes
\be{FloquetSolutionBlowup2}
	c_1^\dagger(\tDagger) \triangleq \int_{t_0^\dagger}^{\tDagger}  
		\vec v_1^T(\tau) \cdot {\vecbphi^\dagger}_{\!2}(\vecphi^*(\tau),\tau) \,\, d\tau.
\ee

\subsection{Choosing $\ah(t)$ to circumvent breakdown of linearization}
To avoid unbounded growth of $\vecdphi^\dagger(\tDagger)$, which would invalidate the present time-shifted
linearization procedure in the same manner as \secref{linearizationInvalid} 
and \lemref{smallDeviationAssumptionInvalid} previously, it is imperative that $c_1^\dagger(\tDagger)$ in
\er{FloquetSolutionBlowup2} remain small and bounded (with respect to ${\vecbphi^\dagger}_{\!2}(\cdot,\cdot)$).
This can be achieved by the simple expedient of requiring that the integrand in \er{FloquetSolutionBlowup2} vanish,
\ie,
\be{C1integrandZero}
	\vec v_1^T(\tau) \cdot {\vecbphi^\dagger}_{\!2}(\vecphi^*(\tau),\tau) \equiv 0, \forall \tau.
\ee
Substituting $\tDagger$ for $\tau$ in \er{C1integrandZero}, and applying \er{bphitwodaggerDefn}, \er{vecbphiTwoDefn},
\er{v1Tdotu1Eq1}, \er{Ktdefn} and \er{tDaggerDefn}, we obtain:
\be{HierPPVequation}
	\begin{split}
	&0  = \vec v_1^T(\tau) \cdot {\vecbphi^\dagger}_{\!2}(\vecphi^*(\tau),\tau) \\
	&\Rightarrow 0  = \vec v_1^T(\tDagger) \cdot {\vecbphi^\dagger}_{\!2}(\vecphi^*(\tDagger),\tDagger) \\
	&\Rightarrow 0  = \vec v_1^T(\tDagger) \cdot \left[ \vecbphi(\phi^*(\tDagger),t) - K(t) \vec u_1(\tDagger)
			\right] \\
	&\Rightarrow K(t) \vec v_1^T(\tDagger) \cdot \vec u_1(\tDagger) = \vec v_1^T(\tDagger)\cdot 
		\vecbphi(\phi^*(\tDagger),t) \\
	&\Rightarrow K(t) = \vec v_1^T(\tDagger)\cdot \vecbphi(\phi^*(\tDagger),t) \\
	&\Rightarrow \dot \ah(t) = \vec v_1^T(\tDagger)\cdot \vecbphi(\phi^*(\tDagger),t) \\
	&\Rightarrow\boxed{\ddt{\ah(t)}{t}=\vec v_1^T(t+\ah(t)) \cdot  \vec b_\phi\big(\vec \phi^*(t+\ah(t)),t\big)}.
	\end{split}
\ee

From the considerations of \secref{TSlinearization} through \eqnref{HierPPVequation}, we are able to prove the
following Theorem: 

\begin{theorem} 
\thmlabel{HierPPVtheorem}
Given a system of $N$ coupled oscillators modelled in the phase domain by the CPS equations
\er{CPS} and mutually injection locked, satisfying \assref{OscGroupIsInjectionLocked} and \assref{OnlyOneFMisOne}.
If the
external perturbations to the system $\{a_i(t)\}$ \er{coupledOscbtphi} (equivalently, $\vecbphi(\cdot,\cdot)$ in
\er{CPS}) are small, and if $\ah(t)$ is chosen to satisfy \er{HierPPVequation}, \ie,
\[
\ddt{\ah(t)}{t}=\vec v_1^T(t+\ah(t)) \cdot  \vec b_\phi\big(\vec \phi^*(t+\ah(t)),t\big),
\]
then the solution of the CPS can be expressed as in \er{TimeShiftedDeviationForm}, \ie, as
\[
	\vec \phi(t) = \vecphi^*\big(t+\ah(t)\big) + {\vecdphi}(t),
\]
where $\vecphi^*(t)$ is the periodic, synchronized solution of the externally-unperturbed system of oscillators.
$\vecdphi(t)$, \textbf{the deviations from the orbit of the externally-unperturbed system,
remain small and bounded for all $t$} (with respect to the external perturbations $\{a_i(t)\}$).
\begin{proof}
Subject to \assref{TimeShiftedDeviationForm} and \assref{KtIsSmall}, \lemref{vecbhi2isSmall} establishes that
${\vecbphi}_{\!2}(\vecphi^*(\tDagger),t)$ is small; applying \er{bphitwodaggerDefn} shows that 
${\vecbphi^\dagger}_{\!2}(\cdot,\cdot)$, which appears in \er{FloquetSolutionOfLPTVtsLinearizedCPS}, is also small.

Choosing $\ah(t)$ to satisfy \er{HierPPVequation} ensures that $c_1^\dagger(\tDagger)$ \er{FloquetSolutionBlowup2}
vanishes, as demonstrated above. As a result, the $i=1$ terms in \er{FloquetSolutionOfLPTVtsLinearizedCPS} (which
correspond to Floquet multiplier $\rho_1=1$ or equivalently, Floquet exponent $\mu_1=1$) remain bounded
and small. From \assref{OnlyOneFMisOne} and \lemref{NoFMisGreaterThanOne}, the remaining Floquet exponents $\mu_2,
\cdots, \mu_N$ all have strictly negative real parts. With ${\vecbphi^\dagger}_{\!2}(\cdot,\cdot)$ small
as noted above, this implies that the terms corresponding to $i=2,\cdots,N$
in \er{FloquetSolutionOfLPTVtsLinearizedCPS} also remain bounded and small for all $t$. As a result, 
$\vecdphi^\dagger(\tDagger)$ remains bounded and small for all $t$. Applying
\er{vecdphidaggerDefn}, $\vecdphi(t)$ also remains small and bounded for all time. This immediately validates
\assref{TimeShiftedDeviationForm}. That $\vecbphi(\cdot,\cdot)$ is small (by assumption) and \er{HierPPVequation}
holds also validates \assref{KtIsSmall}.

Differentiating \er{FloquetSolutionOfLPTVtsLinearizedCPS} and proceeding in a similar manner,
\assref{dvecdphiIsSmall} can also be shown to be valid.

\end{proof}
\end{theorem}

\section{Conclusion}
\er{HierPPVequation} and \thmref{HierPPVtheorem} establish that $\ah(t)$, the time shift (or phase) of the coupled PPV
system, obeys a relationship identical in form to the PPV equation \er{PPVeqn} for individual oscillators. In other
words, groups of synchronized oscillators may be abstracted by a single ``effective PRC/PPV'' function that
dictates the group's ``effective phase response'' $\ah(t)$ to external perturbations via the \textit{single, scalar}
differential equation \er{HierPPVequation}. As such, it provides a rigorous basis for the empirical practice of
measuring PRCs of complex oscillatory systems that are synchronized (\eg, \cite{phylliszeeNSMC2007talk}).

\er{HierPPVequation} may be used to analyze the dynamics (\eg, noise and locking/pulling behaviour) of groups of
synchronized oscillators, just as \er{PPVeqn} is used for individual oscillators.  Moreover, \thmref{HierPPVtheorem}
may be applied repeatedly to abstract the effective PRC/PPV of groups of synchronized oscillators that are organized
hierarchically over multiple levels, enabling the development of computationally efficient and scalable methods for
analysing and abstracting the phase dynamics of large systems of coupled oscillators.

\section*{Acknowledgments}
The author thanks Somil Bansal for pointing out several errors in version 1 of the manuscript.

\let\em=\it

\vfill

\bibliographystyle{unsrt}
\bibliography{all}

\end{document}